%% file: main.tex
\documentclass[10pt,conference]{IEEEtran}
\IEEEoverridecommandlockouts
\usepackage{amsmath,amssymb,amsfonts}
\usepackage{comment}
\usepackage{cite}
\usepackage{graphicx}
\usepackage{textcomp}
\usepackage[breaklinks,colorlinks,linkcolor=blue,citecolor=blue,urlcolor=blue]{hyperref}
\usepackage{colortbl}
\usepackage{multirow}
\usepackage{footnote}
\usepackage{tablefootnote}
\usepackage{float}
\usepackage{threeparttable}
\usepackage{pifont}
\usepackage{textcomp}
\usepackage{tabularx}
\usepackage[table]{xcolor}
\usepackage{placeins}

\usepackage{algorithmic}
\usepackage[ruled,vlined,linesnumbered,commentsnumbered]{algorithm2e}
\usepackage{arydshln}

\def\BibTeX{{\rm B\kern-.05em{\sc i\kern-.025em b}\kern-.08em
    T\kern-.1667em\lower.7ex\hbox{E}\kern-.125emX}}

\begin{document}

\newcommand{\ml}[1]{{\color{red}\bf [Meng: #1]}}
\newcommand{\yx}[1]{{\color{blue}\bf [Yixuan: #1]}}
\newcommand{\zjw}[1]{{\color{green}\bf [zjw: #1]}}
\newcommand{\xt}[1]{{\color{black} #1}}
\newcommand{\xietong}[1]{{\color{purple}\bf [xt: #1]}}

\newcommand{\red}[1]{{\color{red}\bf (#1)}}
\newcommand{\method}{ReaLM}

\title{The Quest for Reliable AI Accelerators: Cross-Layer Evaluation and Design Optimization\vspace{-5pt}}
\author{
    Meng Li$^{123*}$,
    Tong Xie$^{21}$,
    Zuodong Zhang$^{4}$,
    and Runsheng Wang$^{234*}$
\\
\textit{$^1$Institute for Artificial Intelligence \& $^2$School of Integrated Circuits, Peking University, Beijing, China} \\
\textit{$^3$Beijing Advanced Innovation Center for Integrated Circuits, Beijing, China} \\
\textit{$^4$Institute of Electronic Design Automation, Peking University, Wuxi, China} \\
\thanks{This work was supported in part by NSFC (62495102, 92464104, 62125401), National Key Research
and Development Program (2024YFB4505004), Beijing
Municipal Science and Technology Program (Z241100004224015), Beijing Outstanding Young Scientist Program (JWZQ20240101004), and the 111 Project (B18001).}
\thanks{*Corresponding author: \{meng.li,r.wang\}@pku.edu.cn}
\vspace{-20pt}

}

\newcommand{\opt}{\texttt{OPT-1.3B}}
\newcommand{\llama}{\texttt{LLaMA-2-7B}}
\newcommand{\llm}{\texttt{LLaMA-3-8B}}

\maketitle

\input{docs/00_abstract}

\input{docs/01_introduction}
\input{docs/02_avatar}
\input{docs/03_read}
\input{docs/04_realm}
\input{docs/05_results}

\bibliographystyle{ieeetr}

\bibliography{top_simplified.bib,reference_simplified_modified.bib}
\end{document}

%% file: docs/00_abstract.tex
\begin{abstract}
   As the CMOS technology pushes to the nanoscale, aging effects and process variations have become increasingly pronounced, posing significant reliability challenges for AI accelerators. Traditional guardband-based design approaches, which rely on pessimistic timing margin, sacrifice significant performance and computational efficiency, rendering them inadequate for high-performance AI computing demands. Current reliability-aware AI accelerator design faces two core challenges: (1) the lack of systematic cross-layer analysis tools to capture coupling reliability effects across device, circuit, architecture, and application layers; and (2) the fundamental trade-off between conventional reliability optimization and computational efficiency. To address these challenges, this paper systematically presents a series of reliability-aware accelerator designs, encompassing (1) aging and variation-aware dynamic timing analyzer, (2) accelerator dataflow optimization using critical input pattern reduction, and (3) resilience characterization and novel architecture design for large language models (LLMs). By tightly integrating cross-layer reliability modeling and AI workload characteristics, these co-optimization approaches effectively achieve reliable and efficient AI acceleration. 
   
\end{abstract}

%% file: docs/01_introduction.tex
\section{Introduction}

Recently, AI has emerged as a game-changing technology and has revolutionized many different applications, including reliability-critical tasks such as autonomous driving etc. With high computational requirements, AI workloads, particularly large language models (LLMs), have been widely deployed on customized accelerators such as TPU-like systolic arrays. However, as Moore's law pushes to the nanoscale, these accelerators are susceptible to hardware faults. These faults, including timing errors, arise from various reliability issues, such as process, voltage, temperature, and aging (PVTA) variations \cite{huang2017variability, wang2021can, wang2022cross, sun2024challenges}. Therefore, it is crucial to enable both reliable and efficient AI acceleration. 

\begin{figure}
    \centering
    \includegraphics[width=1.0\linewidth]{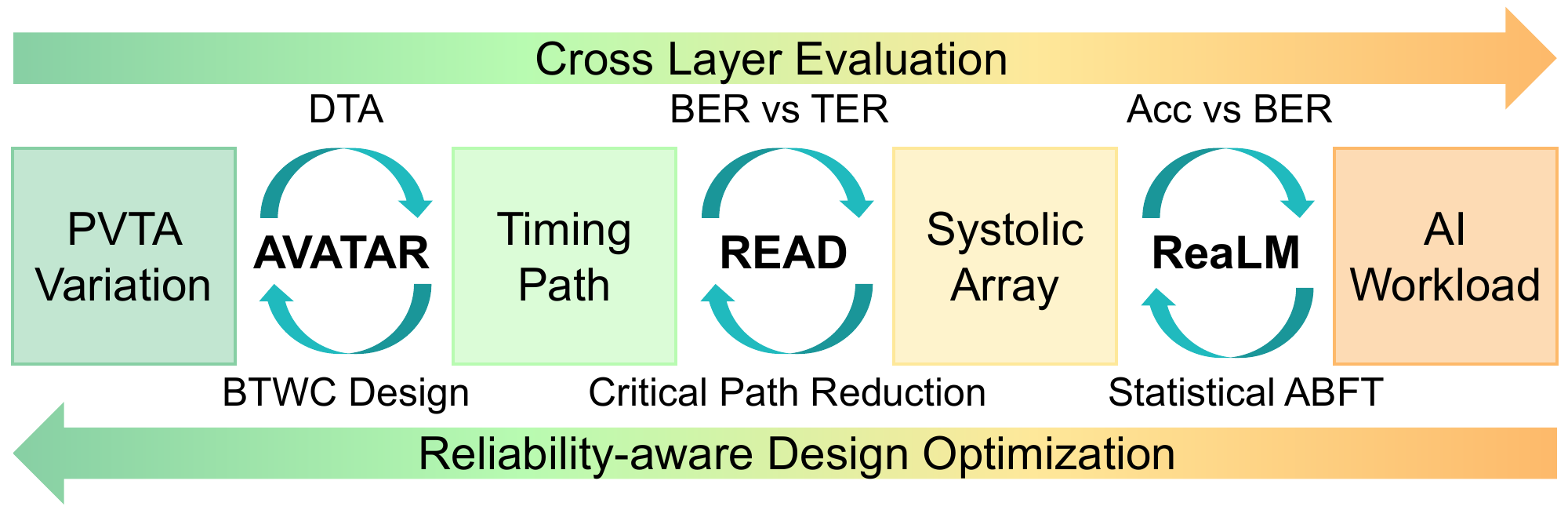}
    \vspace{-15pt}
    \caption{Overview of our recent works.}
    \vspace{-5pt}
    \label{fig:recent_works}
\end{figure}

However, ensuring reliability in AI accelerators presents significant challenges. First, reliability issues exhibit strong cross-layer coupling and lack accurate modeling. For example, at the device level, voltage stress induces threshold-voltage shifts ($\Delta V_{\rm th}$); at the circuit level, these shifts manifest as specific timing-error patterns and timing error rates (TERs); and at the application level, such errors degrade AI model accuracy (Acc). These relationships are highly complex and difficult to model precisely. Second, existing mitigation methods fail to balance reliability and efficiency. Worst-case voltage margins eliminate timing errors but incur heavy overhead, while reduced margins increase bit error rates (BERs) and degrade model performance. Classical techniques such as Razor flip-flops \cite{ernst2003razor} and algorithm-based fault tolerance (ABFT) \cite{huang1984algorithm} are either unscalable to large accelerators or incur frequent yet unnecessary recovery, making them impractical for compute-intensive AI workloads.
To address these challenges, this paper summarize our recent efforts for cross layer evaluation and design optimization techniques \cite{zhang2022avatar, zhang2023read, xie2025realm}, as depicted in Fig. \ref{fig:recent_works}.

%% file: docs/02_avatar.tex
\section{Aging- and Variation-Aware DTA}

\subsection{Motivation}

As technology scaling amplifies timing guardbands, \textit{better-than-worst-case} (BTWC) design seeks to improve efficiency by exploiting dynamic timing margins. Conventional dynamic timing analysis (DTA) tools add worst-case aging and variation guardbands, inflating margins and diminishing BTWC gains. AVATAR~\cite{zhang2022avatar} addresses this by incorporating transistor aging and process variation directly into DTA, eliminating unnecessary guardbands and preserving performance and efficiency.

\vspace{-5pt}
\subsection{AVATAR Algorithm}
\vspace{-3pt}
\begin{figure}[!tb]
    \centering
    \includegraphics[width=0.8\linewidth]{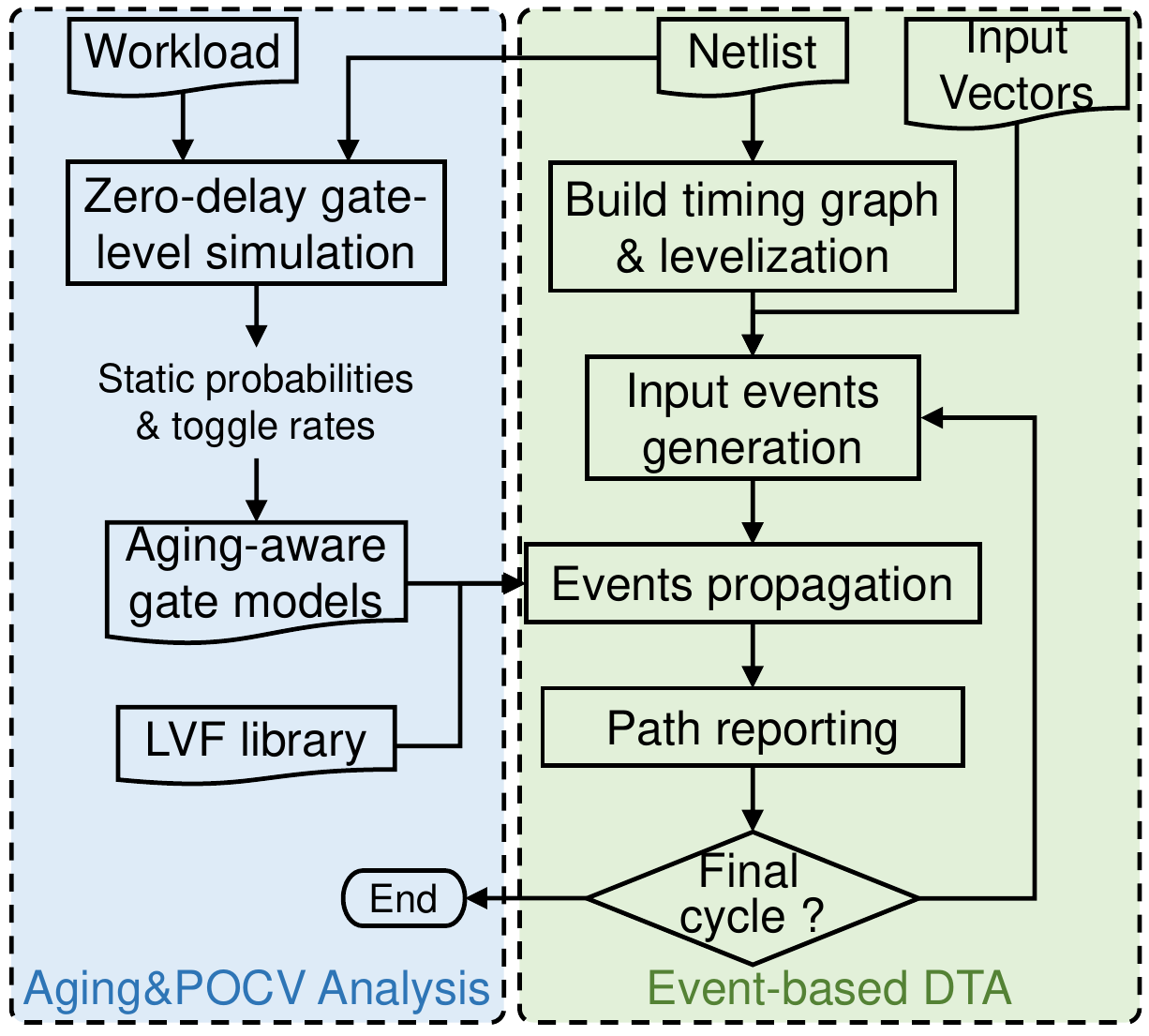}
    \caption{Aging- and variation-aware DTA flow of AVATAR.}
    \label{fig:avatar}
    \vspace{-10pt}
\end{figure}

The core idea of AVATAR is to integrate aging analysis into the DTA process, as illustrated in Fig. \ref{fig:avatar}. The algorithm consists of three main steps: (1) gate-level aging and variation model characterization, (2) workload analysis to estimate $\Delta V_{\rm th}$ for each transistor, and (3) event-based DTA.

In Step 1, the aging model takes as inputs the timing arc, input slew, output load, per-transistor $\Delta V_{\rm th}$, $V_{\mathrm{DD}}$, and temperature, and outputs aged cell delay and transition time using a first-order Taylor expansion. The variation model is based on POCV analysis with LVF data.
In Step 2, zero-delay gate-level simulation is performed to obtain the toggle rate of each internal net, after which the aging model is applied to compute the $\Delta V_{\rm th}$ of each transistor.
In Step 3, events are defined as digital switching activities on pins. Event-based DTA begins by constructing a timing graph for event propagation, followed by cycle-by-cycle analysis that includes input event generation, event propagation, and path reporting, with the arrival times at all timing endpoints recorded.

\subsection{Experiments}

\begin{table}[!tb]
\centering
\caption{Performance improvement from application-based DVFS based on the corner-based DTA and AVATAR}
\label{table:performance}
\resizebox{\linewidth}{!}{
\begin{tabular}{c|cc|cc}
\hline \hline
\multirow{3}{*}{Benchmark} & \multicolumn{2}{c|}{Corner-based DTA~\cite{ISCA_16_DTS, date_2015_instruction-based-clock}} & \multicolumn{2}{c}{AVATAR (this work)} \\ \cline{2-5}
                           & \begin{tabular}[c]{@{}c@{}}Max Freq.\\ (MHz)\end{tabular} & \begin{tabular}[c]{@{}c@{}}Impro. \\ (vs STA)\end{tabular} & \begin{tabular}[c]{@{}c@{}}Max Freq.\\ (MHz)\end{tabular} & \begin{tabular}[c]{@{}c@{}}Impro. \\ (vs STA)\end{tabular} \\ \hline \hline
SHA                        & 948  & 13.75\% & 1020 & 22.38\% \\ \hline
AES\_CBC                   & 883  & 5.99\%  & 951  & 14.10\% \\ \hline
FIR                        & 915  & 9.82\%  & 986  & 18.35\% \\ \hline
BubbleSort                 & 1290 & 55.38\% & 1380 & 65.36\% \\ \hline
Motion\_Detection          & 958  & 15.00\% & 1030 & 23.97\% \\ \hline
CNN                        & 868  & 4.18\%  & 936  & 12.30\% \\ \hline
Convolution                & 868  & 4.19\%  & 936  & 12.28\% \\ \hline
2d\_Filter                 & 936  & 12.33\% & 1050 & 26.37\% \\ \hline
MatrixMult                 & 916  & 9.89\%  & 989  & 18.63\% \\ \hline
DCT                        & 1170 & 40.77\% & 1270 & 52.15\% \\ \hline \hline
\end{tabular}
} 
\vspace{-5pt}
\end{table}

To evaluate the benefits of AVATAR, we determine the application-specific $V_{\min}$/$f_{\max}$ using two methods. (1) Corner-based DTA with extra guardbands \cite{ISCA_16_DTS, date_2015_instruction-based-clock}. Dynamic delay is computed as $delay\times(1+total\_guardband)$, where the aging guardband is assumed to be 15\% and the random variation guardband 5\% at nominal $V_{\mathrm{DD}}$. The trend of guardband versus $V_{\mathrm{DD}}$ is characterized using FO4 delay as a representative cell. (2) AVATAR-based analysis. AVATAR inherently accounts for both aging and variation, eliminating the need for extra guardbands. The final delay is calculated as $\mu(delay)+3\sigma(delay)$. Table \ref{table:performance} shows the performance improvement by applying the maximum frequency at the nominal 0.8V.

%% file: docs/03_read.tex
\section{Critical Input Pattern Reduction}


\subsection{Computing Sequence and Timing Error}

READ \cite{zhang2023read} proposes a reliability-enhanced accelerator dataflow optimization technique that can effectively reduce timing errors.
For a given cycle, timing errors are primarily determined by two factors: (1) the input pattern, which dictates the activated paths, and (2) the operating conditions (e.g., temperature, voltage, aging) that influence the delay of these paths. While operating conditions are often uncontrollable, only a small subset of paths, i.e., critical paths, typically exceed the clock period. Thus, reducing the activation rate of these critical paths is key to lowering the timing error rate.

\begin{figure}[!tb]
    \centering
    \includegraphics[width=0.7\linewidth]{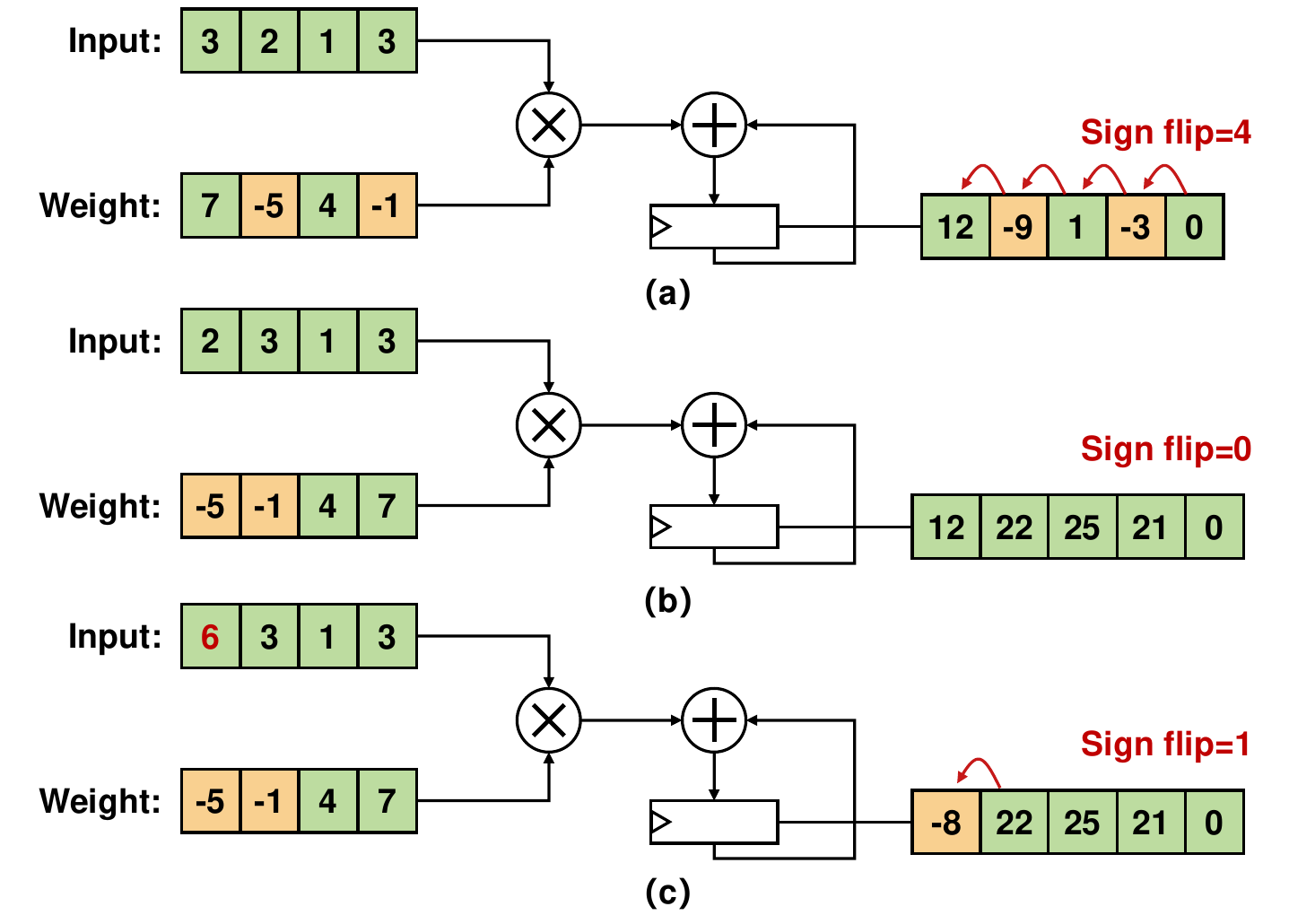}
    \caption{A 1$\times$4 convolution calculated in different orders. Reordering weights does not change the computing result, but avoids the critical input pattern.}
    \label{fig:reorder-schematic}
    \vspace{-10pt}
\end{figure}

The basic processing element (PE) in our study is a multiplier-and-accumulator (MAC) unit. Using the 8-bit multiplier and 24-bit accumulator of a TPU MAC as an example, dynamic timing analysis of the synthesized unit reveals that the most frequent critical input patterns are those that flip the sign bit of the partial sum (PSUM). For instance, computing $3\times(-2)+2=-4$ produces the 2’s complement encoding $111111111111111111111100$, where the sign bit flip triggers a long carry chain in the accumulator, activating critical paths. A similar effect occurs when PSUM transitions from negative to positive.
Consequently, reducing the PSUM sign-flip rate is an effective strategy for minimizing timing error rates.

\subsection{Input Channel Reordering and Output Channel Clustering}

\begin{figure}
    \centering
    \includegraphics[width=0.9\linewidth]{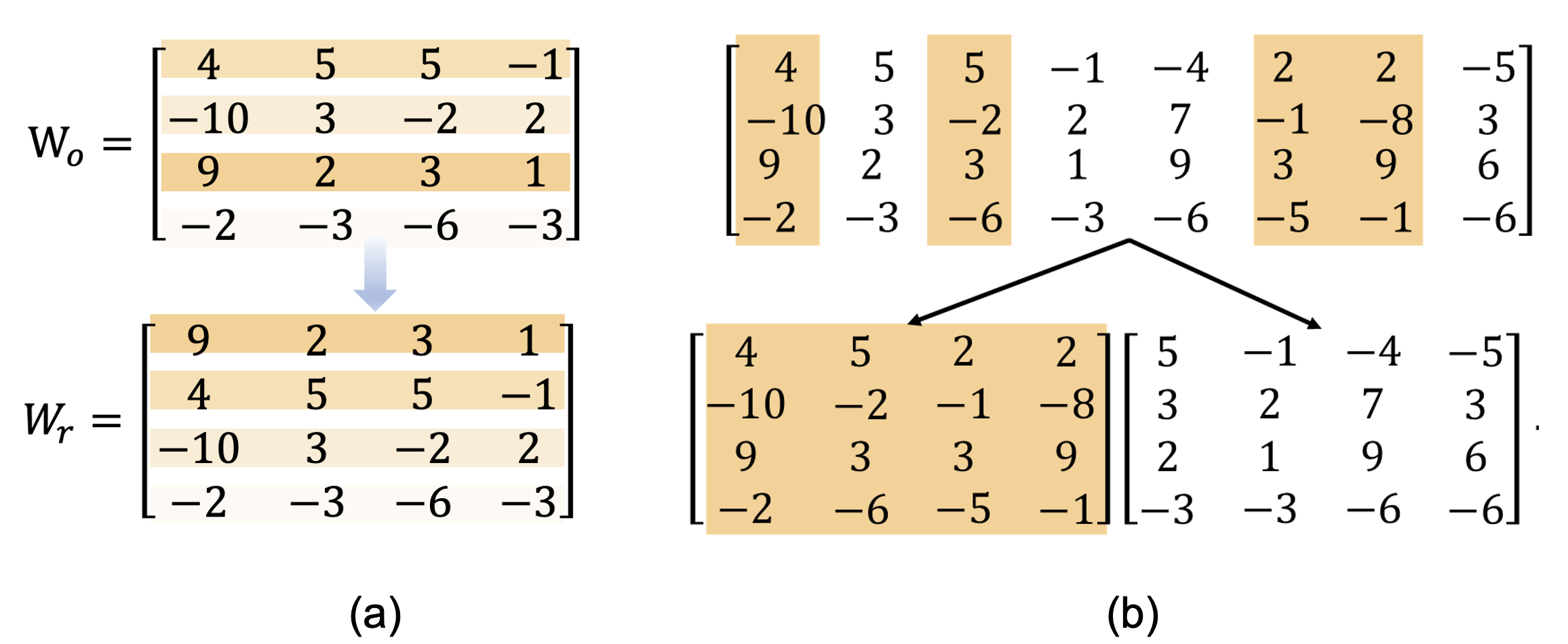}
    \vspace{-10pt}
    \caption{(a) Input channel reordering. (b) Output channel clustering.}
    \vspace{-10pt}
    \label{fig:read_method}
\end{figure}

Given that the ReLU function predominantly results in non-negative outcomes, our heuristic prioritizes non-negative weight computations, as shown in Fig. \ref{fig:read_method} (a). This is achieved via a channel-wise reordering algorithm that sorts operations by the fraction of positive weights within the same channel.



To enhance effectiveness when the number of columns $A_c$ is large, we adopt a \textit{cluster-then-reorder} strategy: output channels are first clustered, and then input channels within each cluster are reordered. As illustrated in Fig.~\ref{fig:read_method} (b), instead of directly segmenting $W$, we first cluster the output channels into groups and then segment $W$ into two sub-matrices. Clustering groups channels with similar sign patterns facilitates input-channel reordering within each sub-matrix.

Formally, we define the \textbf{sign difference (SD)} between two $n$-dimensional vectors $\mathbf{x}$ and $\mathbf{y}$ as the Manhattan distance of their sign vectors.
Minimizing the SD within each sub-matrix simplifies input-channel reordering and reduces sign bit flips. The output-channel clustering problem is thus formulated as minimizing $\mathrm{SD}(W_{T_i})$ across all clusters, and we adopt the balanced KNN on the weight sign matrix by the Manhattan metric to solve this problem.


\vspace{-2pt}
\subsection{Experiments}
\vspace{-2pt}

\begin{figure}[tb]
    \centering
    \includegraphics[width=1.0\linewidth]{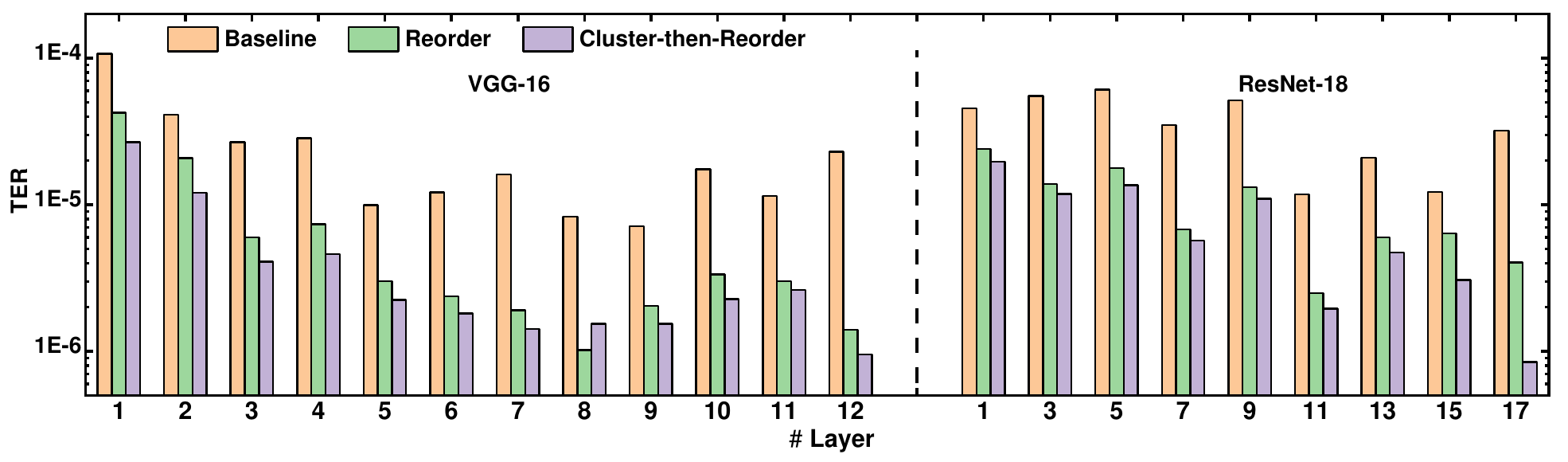}
    \vspace{-15pt}
    \caption{Timing error rate comparison for different reliability-enhanced algorithms on ResNet-18 and VGG-16.}
    \vspace{-5pt}
    \label{fig:ter-layer}
\end{figure}

As shown in Fig. \ref{fig:ter-layer}, the direct reordering and cluster-then-reorder algorithms achieve average TER reductions of $4.9\times$ and $7.8\times$, respectively. The cluster-then-reorder approach consistently delivers greater reductions across most layers and exhibits superior performance in later network layers, where the number of output channels is larger.

%% file: docs/04_realm.tex
\section{Statistical Algorithm-based Fault Tolerance}

\subsection{LLM Resilience Characterization}

\begin{figure*}[!tb]
    \centering
    \includegraphics[width=1\linewidth]{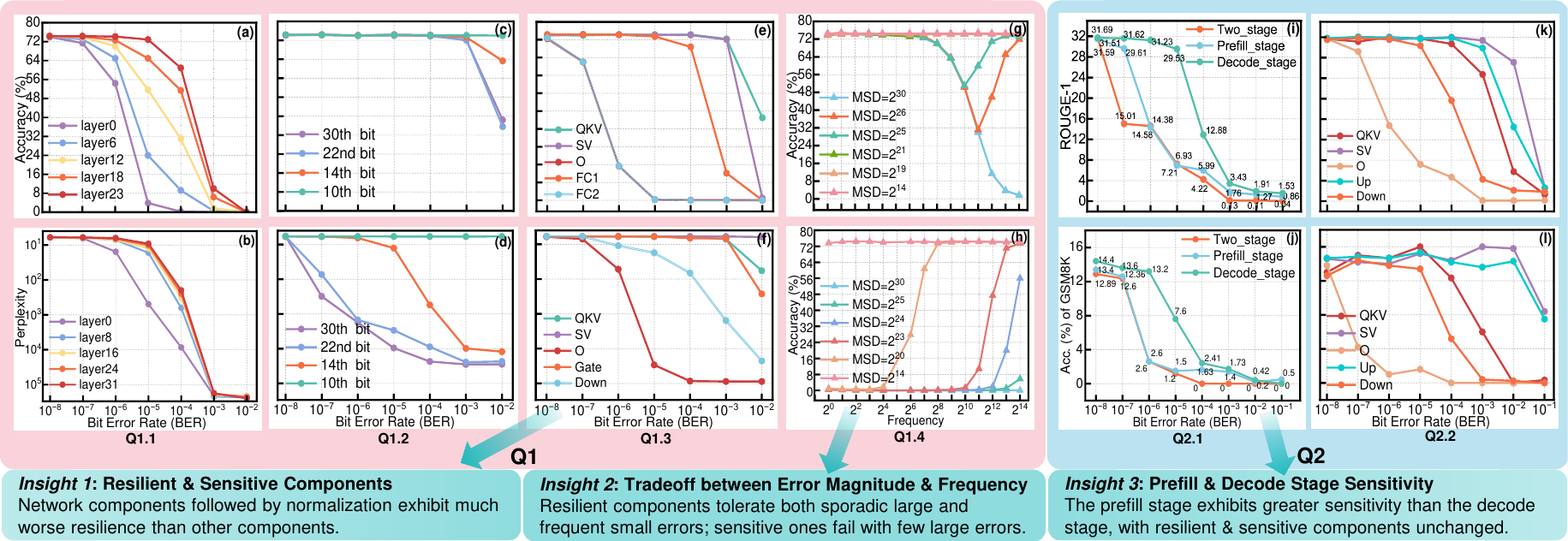}
    \vspace{-15pt}
    \caption{\textbf{Q1.1}: (a)(b) Layer-wise resilience of different LLMs on different tasks. 
    \textbf{Q1.2}: Bit-wise error resilience. (c) Error injection on \texttt{K}. (d) Error injection on \texttt{O}.  
    \textbf{Q1.3}: (e)(f) Sensitivity to errors in different LLM components. 
    \textbf{Q1.4}: Relationship between error frequency and magnitude. (g)  Resilient components like \texttt{K}. (h) Sensitive components like \texttt{O} Given MSD, the error magnitude decreases as the error frequency increases. 
    \textbf{Q2.1}: (i)(j) Comparison between the prefill stage and decode stage.
    \textbf{Q2.2}: (k)(l) Impact of error injection across network components: \texttt{O} and \texttt{Down} remain highly sensitive. 
    (a)(c)(e)(g)(h) are evaluated with \opt~on LAMDABA; (b)(d)(f) with \llama~on WikiText-2; (i)(k) with \llama~ on X-Sum; (j)(l) with \llama~on GSM8K.     
    }
    \label{fig:resilience characterization}
    \vspace{-15pt}
\end{figure*}

ReaLM \cite{xie2025realm} characterizes the error resilience of LLMs and proposes statistical ABFT to adaptively corrects only those critical errors.
In this section, we aim to answer six questions.  
\textbf{Q1.1}: How does resilience vary across different layers of LLMs?  \textbf{Q1.2}: What is the bit-wise resilience of LLMs? 
\textbf{Q1.3}: How does resilience vary among different components within LLMs during the prefill stage? 
\textbf{Q1.4}: What is the correlation between fault magnitude and frequency in impacting LLM performance? 
\textbf{Q2.1}: How does the resilience of LLMs compare between the prefill and decode stages?
\textbf{Q2.2}: How does resilience vary among different computational components within LLMs during the decode stage? 

Fig. \ref{fig:resilience characterization} illustrates the resilience characterization for these questions. For \textbf{Q1.1}, \textit{while the layers exhibit comparable resilience behaviors, the earlier layers are more vulnerable to errors.} For \textbf{Q1.2}, we conclude that \textit{while errors at lower bits have a negligible impact on model performance, errors at higher bits can reach a saturation point due to re-quantization.}

For \textbf{Q1.3}, we find that \textit{network components followed by normalization operations exhibit much worse resilience}. On the other hand, components like \texttt{QKV} are much more resilient. So, we can divide these components into two categories: the sensitive components, such as \texttt{O} and \texttt{Down} projection, and the resilient components, such as \texttt{QKV}. 

\textbf{Q1.4} is about the error magnitude and frequency.
As depicted in Fig. \ref{fig:resilience characterization} (g)(h), we find that given a total sum of errors, as the error frequency increases and error magnitude decreases, the resilient components exhibit a non-monotonic resilience behavior. \textit{Resilient components can tolerate both sporadic large and frequent small errors. But for sensitive components, even a few large errors can severely degrade performance}. This finding aligns with our earlier observations.

For \textbf{Q2.1} and \textbf{Q2.2},
Fig. \ref{fig:resilience characterization} (i) and (j) reveal that \textit{the prefill stage exhibits greater sensitivity to errors than the decode stage}. 
And Fig. \ref{fig:resilience characterization} (k) and (l) further confirm that \textit{the resilient and sensitive components remain consistent across both stages}.

\vspace{-2pt}
\subsection{Statistical Algorithm-Based Fault Tolerance}
\vspace{-2pt}

\begin{figure}[!tb]
    \centering    \includegraphics[width=0.9\linewidth]{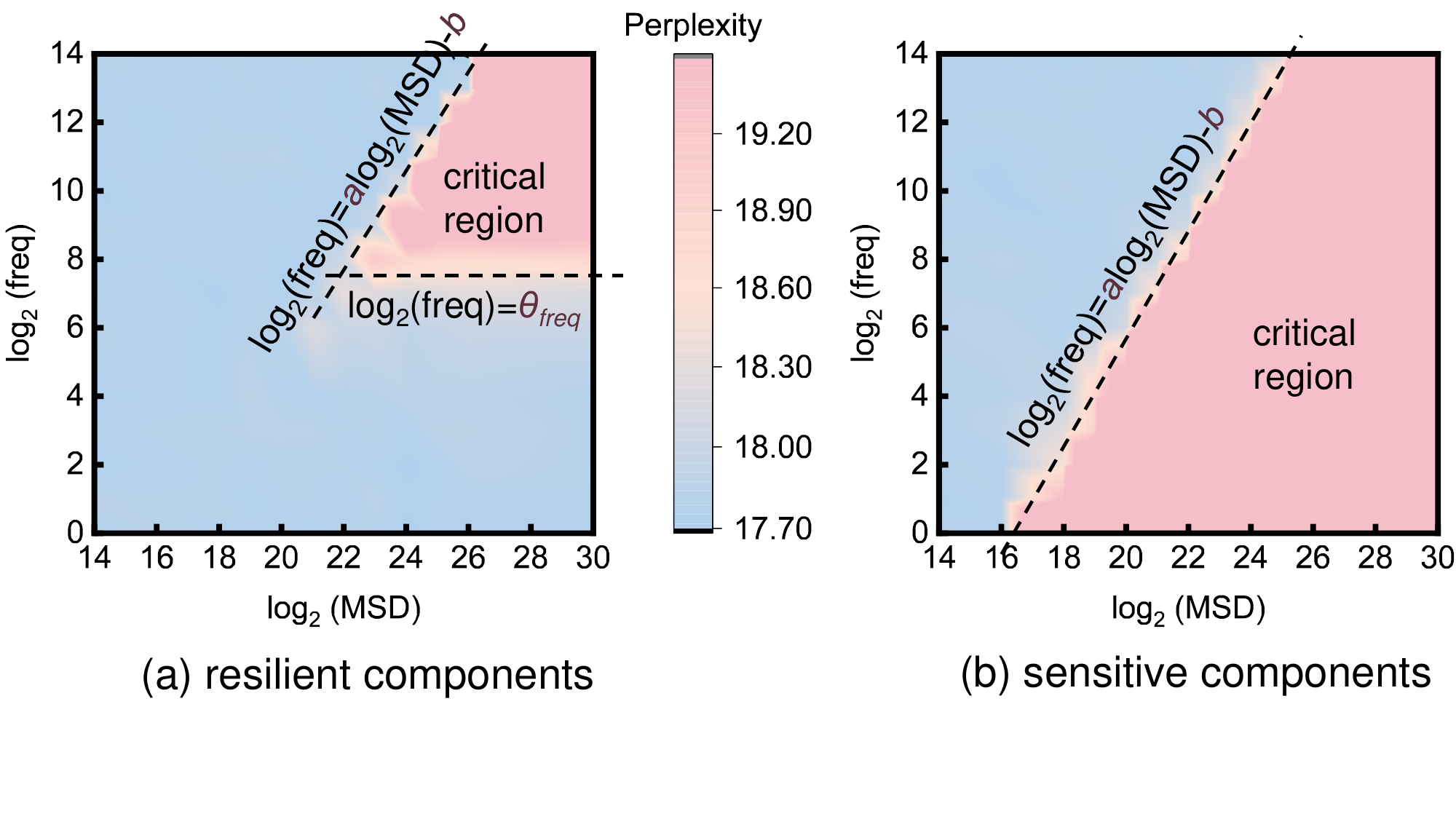}
    \vspace{-22pt}
    \caption{Our statistical ABFT strategy only corrects errors falling inside the critical region.}
    \vspace{-10pt}
    \label{fig:asicon_abft_strategy}
\end{figure}


Our statistically based error detection strategy aims to correct only those critical errors.
Fig. \ref{fig:asicon_abft_strategy} rephrases the tradeoff between error magnitude and frequency.
Given a specific task, we first define an acceptable performance degradation threshold. This threshold defines a critical region where computational errors significantly impact LLM performance. 
The recomputation process is only triggered if the observed errors fall inside the critical region, thereby avoiding unnecessary error recovery process.

\begin{figure}[!tb]
    \centering
    \includegraphics[width=0.85\linewidth]{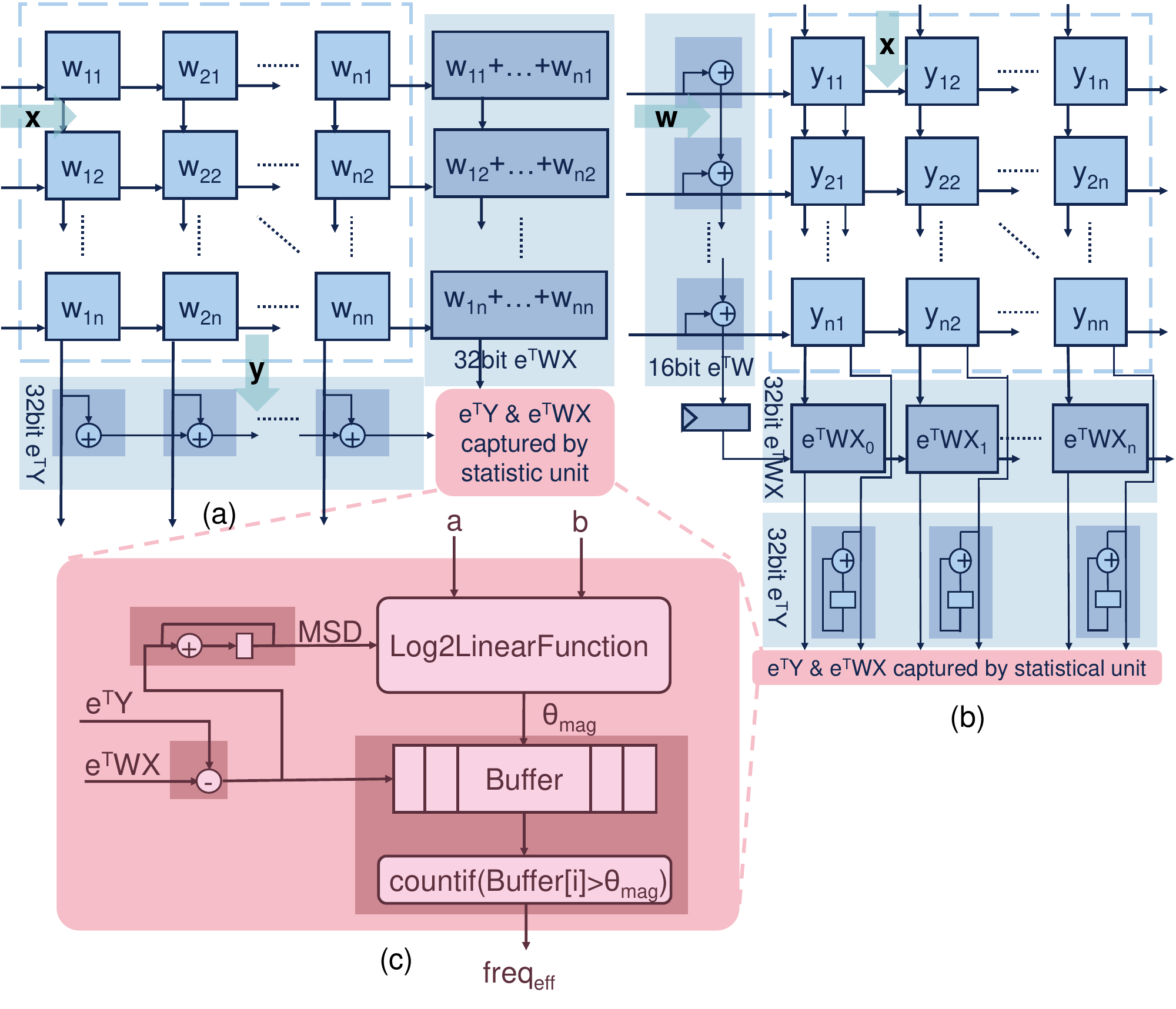}
    \vspace{-10pt}
    \caption{Architecture design of statistical ABFT on systolic array: (a) ABFT implementation for weight stationary dataflow; (b) ABFT implementation for output stationary dataflow; and (c) customized statistical units.}
    \label{fig:architecure_design_asicon}
    \vspace{-20pt}
\end{figure}

Fig. \ref{fig:architecure_design_asicon} illustrates the design of our statistical ABFT circuit. We first take the weight stationary dataflow as an example. A column of PEs is added on the right, storing weight checksums for computing $e^{\rm T}WX$. Meanwhile, a row of adders is appended at the bottom, which accumulates the output checksum. In addition, we introduce a statistical unit to capture the error statistics and determine the need for recovery. Once the observed errors fall inside the critical region, the recomputation process will be triggered.
We also support the output stationary dataflow. A column of adders is appended to the left side to compute the weight checksum, and a row of PEs is added at the bottom to calculate the output checksum.

\vspace{-3pt}
\subsection{Experiments}
\vspace{-3pt}

\begin{figure}[!tb]
    \centering
    \includegraphics[width=1\linewidth]{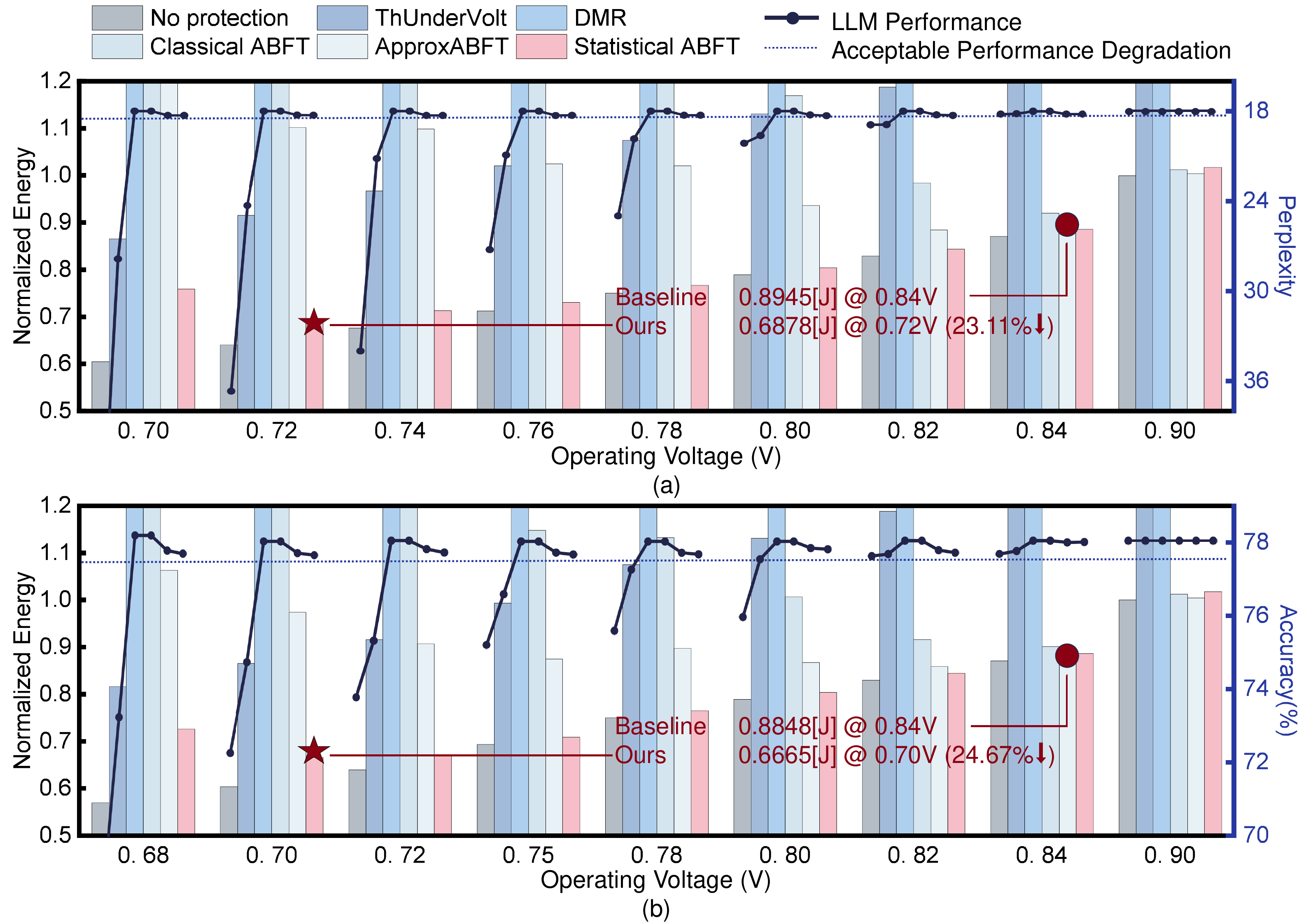}
    \vspace{-20pt}
    \caption{\xt{LLM performance and total energy savings comparison. (a) \opt~on WikiText-2. (b) \llm~on HellaSwag.
    Our method achieves competitive performance with minimal protection overhead, significantly reducing costs versus existing methods, with maximal energy savings at 0.72V and 0.70V, respectively. The dash line marks the acceptable performance degradation.
    }}   
    \label{fig: mainresults_asicon}
    \vspace{-15pt}
\end{figure}

For circuit overhead comparison, compared to the unprotected systolic array, our design introduces only about 1.4\% area and 1.8\% power overhead, making it lightweight and practical for real-world deployment.
We then evaluate the LLM performance and energy savings in Fig. \ref{fig: mainresults_asicon}. We take \texttt{K} in \texttt{OPT-1.3B} on WikiText-2 and \texttt{V} projection in \texttt{LLaMA-3-8B} on HellaSwag as examples. For the \texttt{OPT-1.3B}, our design achieves a sweet point at 0.72 V and saves 23\% energy compared to prior-art methods. And for \texttt{LLaMA-3-8B}, it finds a sweet point at 0.70 V and saves 24\% energy.

%% file: docs/05_results.tex
\section{Conclusion}

Reliability issues in AI accelerators arise from cross-layer coupling across the device, circuit, architecture, and application levels. Accurate reliability modeling can significantly reduce overly conservative design margins, while cross-layer co-optimization effectively mitigates reliability constraints. Many hardware faults may be masked, and AI applications are inherently fault-tolerant, eliminating the need to design for worst-case conditions. Allowing limited timing violations thus opens opportunities for substantial improvements in power, performance, and area.